\documentclass[twocolumn]{pasj00}

\begin{document}

\SetRunningHead{Mrk509}{H. Noda et al.}
\title{Suzaku Studies of Wide-Band Spectral Variability\\
     of the Bright Type I Seyfert Galaxy Markarian 509}

\author{Hirofumi \textsc{Noda},\altaffilmark{1}
Kazuo \textsc{Makishima},\altaffilmark{1}
Shin'ya \textsc{Yamada},\altaffilmark{1}
Shunsuke \textsc{Torii},\altaffilmark{1}\\
Soki \textsc{Sakurai},\altaffilmark{1}
and Kazuhiro \textsc{Nakazawa}\altaffilmark{1}
}
\altaffiltext{1}{
  Department of Physics, University of Tokyo\\
  7-3-1, Hongo, Bunkyo-ku, Tokyo, 113-0033, Japan}

\email{noda@juno.phys.s.u-tokyo.ac.jp}

\Received{2011 May 28}
\Accepted{2011 September 2}
\Published{$\langle$publication date$\rangle$}

\maketitle

\begin{abstract}
The Type I Seyfert galaxy Markarian 509 
was observed with Suzaku in 2010 November, 
for a gross time span of 2.2 days.
Timing and spectral properties of the 0.5--45 keV X-rays, 
detected with the XIS and HXD,
consistently revealed the presence of a soft  spectral component 
that remained constant while the total X-ray intensity varied by $\pm 10$\%.
This stable soft component, found in the 0.5--3.0 keV range,
was  interpreted as a result of thermal Comptonization  in a corona 
with  a temperature of $\sim 0.5$ keV and an optical depth of $\sim 18$. 
The time-avearged  0.5--45 keV Suzaku spectrum was reproduced successfully,
as a combination of this thermal Comptonization component,
a harder power-law of photon index $\sim 1.8$,
moderate reflection, and an iron K-emission line.
By analyzing  four archival Suzaku datasets of the same object obtained  in 2006, 
the thermal Comptonization component,
which was  stable during the 2.2 day pointing in 2010,
was found to vary on time scales of a few weeks,
independently of the  power-law component.
Implications of these results are discussed 
in terms of the ``multi-zone Comptonization'' view,
obtained with Suzaku from the black hole binary Cygnus X-1.

\end{abstract}

\section{Introduction}
\label{sec:intro}

Active Galactic Nuclei (AGNs), namely,
accreting massive  black holes (BHs) residing at the center of some galaxies,
often exhibit a phenomenon called ``soft excess"
(e.g., Arnaud et al. 1985, Turner and Pounds 1989),
which appears  in their soft X-ray spectra 
as a steep flux upturn towards lower energies.
For years, this structure has remained unaccounted,
and a number of explanations have been proposed.
The simplest way to interpret it is a direct black body emission 
from the accretion disk around the central black hole.
However, this interpretation was not considered reasonable,
since this would require an  accretion disk with a temperature of $\sim 2 \times 10^5$~K,
which would be too high for BHs of  $\gtrsim 10^8~M_\odot$
where $M_\odot$ is the solar mass (Gierlinsk and Done 2004).

Other interpretations of the soft excess phenomenon include;
flux leakage through a patchy absorber (e.g., Weaver et al. 1995), 
continuum modification due to  warm absorbers
 (e.g., Schurch \& Done 2008; O'Neil et al. 2007),
and reflection from a photoionized accretion disk 
emitting relativistically blurred Fe-L lines  (Ross, Fabian, and Ballantyne 2002; 
Crummy et al. 2006; Zoghbi et al. 2009; Nardini et al. 2010).
As often argued (e.g, Sobolewska \& Done 2007),
these alternative models degenerate,
and are difficult to observationally distinguish.
We do not even know
whether the soft excess is a homogeneous phenomenon,
or a mixture of effects from different origins.

One of the causes of the above degeneracy is the lack of sufficient 
understanding of the primary continuum emission from these AGNs.
Due mainly to limitations in  broad-band sensitivity of the available data, 
X-ray continua in these AGNs have usually been 
approximated by a single power-law (PL) model,
against which a soft excess is defined.
However, recent Suzaku observations 
of the leading black-hole binary (BHB), Cyg X-1, 
have revealed that its continuum in the Low/Hard state 
is considerably more convex than a single PL shape, 
requiring at least two thermal Comptonized components with different $y$ parameters.
Then, AGN may also exhibit such ``multi-zone Comptonization" (MZC) effects, 
because AGNs and BHBs are generally 
believed to exhibit common accretion phenomena.
Such intrinsic curvature in the X-ray continuum,
if found in AGNs, is expected to significantly renew our
understanding of their soft excess issue.

Because of its genuine broad-band sensitivity,
Suzaku (Mitsuda et al. 2007) obviously provides  
the best available mean to search AGN spectra
for the expected MZC effects.
Given these, we studied the archival Suzaku data of the Seyfert galaxy MCG--6-30-15, 
and successfully detected a hard X-ray component 
that varies independently of the dominant PL continuum (Noda et al. 2011). 
This new component can be interpreted as 
a thermal Compton emission with a relatively high optical depth. 
These recent achievements urge us to examine X-ray continua 
of other Seyfert galaxies for further MZC effects: 
the soft excess phenomenon itself is an obvious candidate.
At least in some cases, 
the soft excess could represent 
an intrinsic continuum curvature due to an MZC effect.
Actually, prior to the Suzaku studies,
Marshall et al. (2003) had already applied 
a similar view to a narrow-line Seyfert I galaxy,
and Porquet et al. (2004) to PG quasars.

To examine the soft excess phenomena for such an MZC interpretation,
we selected the  Type I Seyfert galaxy Markarian 509
(hereafter Mrk 509) 
and observed it with Suzaku for  2010 Novermber 21--23.
This object is suited to our purpose, 
because it is X-ray bright (2-10 keV flux of $\sim2-5 \times 10^{-11}$ erg cm$^{-2}$ s$^{-1}$), 
with a low Galactic hydrogen column density 
of $N_{\rm H}=4.4 \times 10^{20}$ cm$^{-2}$ (Dickey \& Lockman 1990).
This AGN has an estimated BH mass of $1.4 \times 10^8~M_{\odot}$ (Peterson et al. 2004),
and a mass accretion rate of $\sim 0.1$ times the Eddington rate as 
estimated from the optical (510 nm) luminosity in Peterson et al. (2004) 
after applying a bolometric correction by a factor of 9 (Kaspi et al. 2000).
We hence regard the object as in a state analogous to the Low/Hard state of BHBs.
(Although this accretion rate could be a little too high for this analogy to hold, 
state transitions of AGNs are much less understood than those of BHBs.)
To be as model independent as possible,
we conducted combined time-domain and spectrum-domain analysis.
As a result, we have detected, over 0.5--3.0 keV of the spectrum,
a stable soft X-ray excess component,
and successfully interpreted it in terms of the MZC view.

Prior to the present observation, 
Mrk 509 had been observed with Suzaku on 4 occasions in 2006;  
namely, on April 25, October 14, November 15, and November 21, 
for a typical exposure of 25 ks each. 
The obtained data were already utilized by Ponti et al. (2009). 
Although these observations have relatively short exposures, 
and hence not suited to the study of intra-observation variations, 
they provide important insight into long-term spectral changes. 
We hence analyzed these archival data as well, 
and found that the soft excess component varied on longer time scales.

\section{Observation and Data Reduction}
\label{sec:obs}

\begin{figure*}[t]
 \begin{center}
  \FigureFile (80mm,80mm)
    {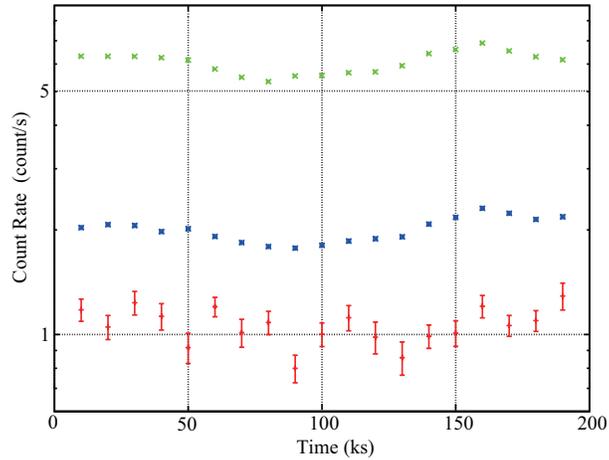}
 \end{center}
\setlength{\belowcaptionskip}{0mm}
 \caption
{Background-subtracted and dead-time corrected
light curves of Mrk 509,
measured with  XIS 0 plus 3 (0.5--3.0 keV in green, 3--10 keV in  blue)
and HXD-PIN (15--45 keV in red),
shown with a binning of 10 ks.
Error bars represent statistical $1\sigma$ ranges.
The HXD-PIN count rates and errors are displayed after multiplied
 by a factor of 10. }
\label{fig:lc}
\end{figure*}

During the Suzaku AO-5 cycle, we observed Mrk 509 for and a net exposure of  102 ks.
The observation started on 2010 November 21 13:59 UT,
lasted for 2.2 days, 
and ended on November 23 19:34 UT.
The source was placed at the XIS nominal position.
Suzaku has four sets of X-ray telescopes (Serlemitsos et al. 2007),
four X-ray CCD cameras called the XIS (X-ray imaging Spectrometer: Koyama et al. 2007),
and the Hard X-ray Detector (HXD: Takahashi et al. 2007; Kokubun et al. 2007).
Among the four XIS sensors, 
XIS 0, 2, and 3 use front-illuminated (FI) CCDs which have almost identical responses,
while  XIS 1 employs a back-illuminated (BI) CCD.
They cover an  energy range of 0.5--10 keV.
The HXD covers a 10--70 keV energy band with Si PIN photo-diodes (HXD-PIN),
and a 50--600 keV range with GSO scintillation counters (HXD-GSO). 
In the present work,
we utilized the XIS and HXD-PIN data, which were prepared via version 2.5 processing.
The available net exposure is 102 ks for the XIS, and 92 ks for the HXD.

On-source events of each CCD camera
were extracted from a circular region of $180''$ radius centered on
the source.
Background events were taken from a surrounding annular region 
of the same camera with the inner and outer radii of $200''$ and of $269''$, respectively.
The response matrices and ancillary response files were 
created by \texttt{xisrmfgen} and \texttt{xissimarfgen} (Ishisaki et al. 2007), respectively.
In the present analysis,
the data of XIS 0 and 3 were added and utilized as XIS FI,
while those of XIS 1 or 2 were not. 

Events of HXD-PIN were processed in a similar way.
Since the HXD has no imaging capability,
Non X-ray Background (NXB) in the data  is estimated by analyzing a set of  fake events
which were created by a standard NXB model (Fukazawa et al. 2009).
We analyzed the on-source events and the NXB events in the same manner,
and subtracted the latter from the former.
In addition, the on-source data includes the Cosmic X-ray Background (CXB; Bolt et al. 1987),
so we need to estimate and  subtract it as well.
The CXB contribution was estimated using the HXD-PIN response to diffuse sources, 
assuming the spectral CXB surface brightness model
determined by HEAO 1 (Gruber et al. 1999):
$9.0 \times 10 ^{-9} (E/3~\mathrm{keV})^{-0.29} \exp (-E/40~\mathrm{keV})$
erg cm$^{-2}$ s$^{-1}$ str $^{-1}$ keV$^{-1}$,
where $E$ is the photon energy.
The estimated CXB count rate is 5\% of the NXB signals.
   
In the same way as described above, we also processed the 2006  Suzaku data
of Mrk 509. The  exposures obtained with the XIS/HXD were
25/15, 26/23, 24/18, and 33/29 ks,  on April 25, October 14,
November 15, and November 27,  respectively.

\section{Analysis of the 2010 Data}

\subsection{Timing analysis}

To look for  time  variations in different energies,
light curves were extracted in three bands of the 2010 data; 
0.5--3.0 keV (soft band),  3.0--10 keV (middle band), and 18.0--45.0 keV (hard band).
Figure 1 shows the derived three light curves with 10 ks binning.
The count rates in the soft and middle bands thus varied by $\pm10$\%, in a good correlation.
Although the HXD-PIN light curve (hard band) 
exhibits statistically significant variations with $\chi^{2}$/d.o.f.$ =31.42/17$,
we cannot tell immediately whether its variation follows those of the other bands.

\begin{figure*}[t]
 \begin{center}
   \FigureFile(120mm,120mm)
    {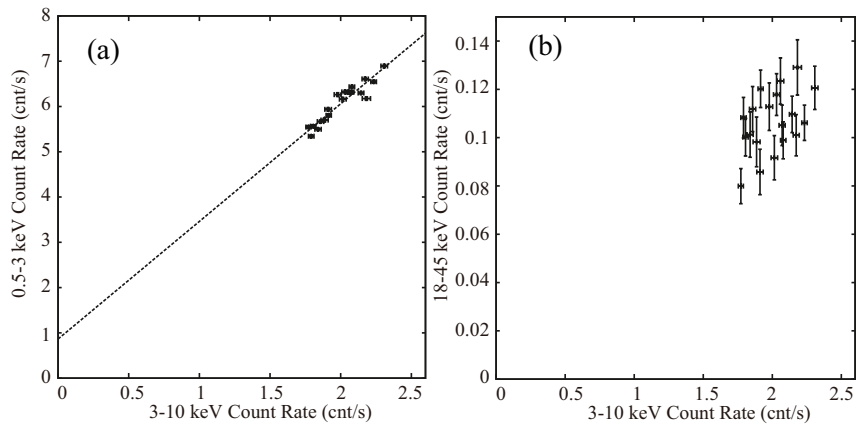}
 \end{center}
\setlength{\belowcaptionskip}{0mm}
 \caption
{Count-Count Plots of Mrk 509 obtained on 2010 november 21--23.
Abscissas gives NXB-subtracted XIS 0 plus 3 count rate in 3--10 keV,
while ordinate gives (a) the XIS 0 + 3 (0.3--3.0 keV) count rate, 
and  (b) those of the HXD-PIN (15--45 keV)
after the NXB subtraction. 
All data are binned into 10 ks.
The dotted straight line in (a) refers to equation (1). }
\label{fig:ccp}
\end{figure*}

To study correlations among the three light curves,
we utilize Count-Count Plots (CCPs), 
in which abscissa is the count rate in the middle band,
and ordinate is those in  the soft or hard bands.  
As expected, the middle vs. soft CCP, shown  in figure 2(a), reveals 
a strong correlation between these two bands.
When abscissa and ordinate are represented by $x$ and $y$, respectively, 
the data distribution in this CCP can be reproduced by a linear relation as,
\begin{eqnarray}
&y= (2.60 \pm0.19)x + (0.87 \pm 0.37),
\label{eq:ccpfit12}
\end{eqnarray}
with $\chi^2 /$d.o.f.$ = 24.33 / 17$. 
Here, a systematic error of 2\% was added to each XIS count, 
which comes from uncertainties in  CCD contamination, 
varying because of the source motion on the detectors by attitude jittering.
Similarly,
the middle vs. hard CCP, shown in figure 2(b), 
represents generally  
positive correlations,
but the data scatter is too large for a linear fit.
The scatter suggests the presence of 
a variable hard component independent of the dominant power-law, 
as reported in MCG--6-30-15 (Noda et al. 2011).
This issue will be studied elsewhere.

\begin{figure*}[t]
 \begin{center}
   \FigureFile(110mm,110mm)
    {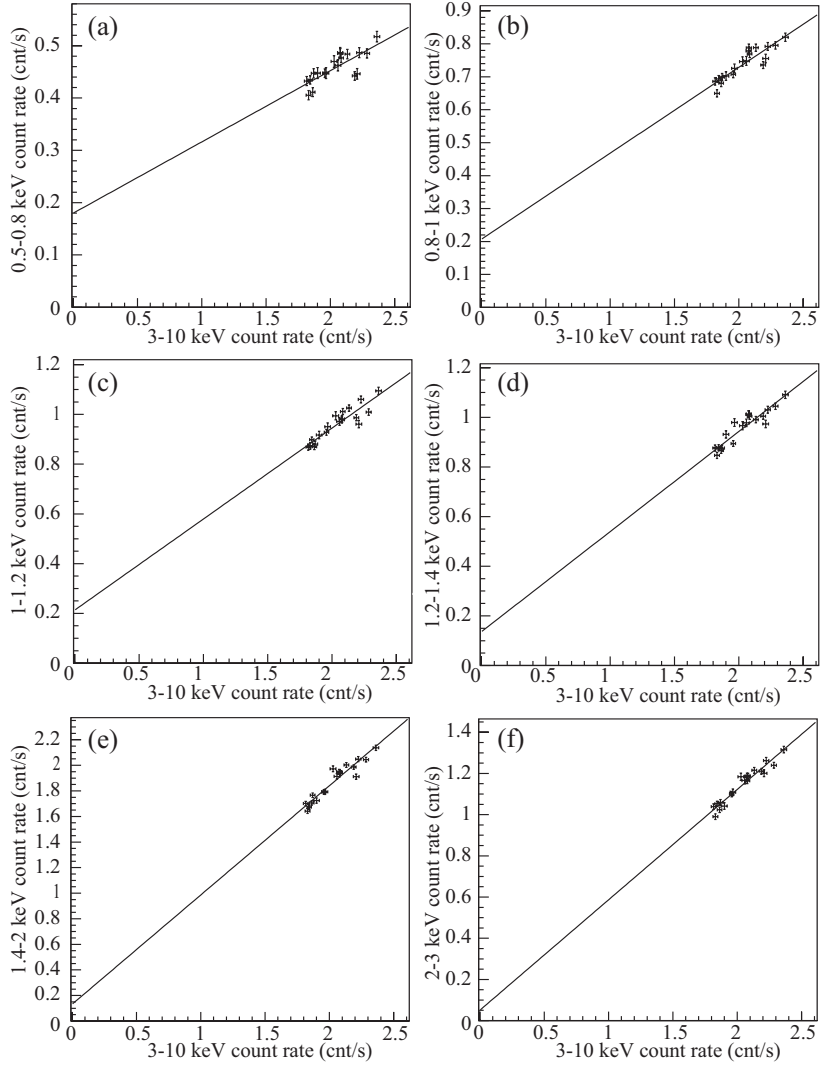}
 \end{center}
\setlength{\belowcaptionskip}{0mm}
 \caption
{The same as figure 2, but calculated in 6 finer soft energy bands
against the 3--10 keV count rate. Ordinate refers to the NXB-subtracted
XIS 0 + 3 count rate in (a) 0.5--0.8 keV,
(b) 0.8--1.0 keV, (c) 1.0--1.2 keV,
(d) 1.2--1.4 keV, (e) 1.4--2.0 keV,
and (f) 2.0--3.0 keV. }
\label{fig:ccp}
\end{figure*}

The above fit to the middle vs. soft  CCP  shows a significant zero-point offset 
(non-zero $y$-intercept).
As already utilized by Churazov et al. (2001) to extract a stable disk emission component 
in the soft state of Cyg X-1, 
this offset in figure 2(a) suggests the presence of a stable emission in the soft band.
Then, to investigate detailed spectral dependence of this effect,
we divided the 0.5--3.0 keV band into 6 finer energy bands,
0.5--0.8 keV, 0.8--1.0 keV, 1.0--1.2 keV, 1.2--1.4 keV, 1.4--2.0 keV, and 2.0--3.0 keV,
and computed a CCP in each band against the 3--10 keV (middle band) count rate.
The derived 6 CCPs are presented in figure 3.
In every CCP,
the data distribute almost linearly with small scatter,
so the correlation between each finer energy band 
and the fiducial  3--10 keV band is considered to be strong.
As a result,
every CCP can be reproduced by a linear relations as
\begin{eqnarray}
y = a x + b,
\label{eq:ccpfit12}
\end{eqnarray}
where $x$ is again the middle-band count rate, while $y$ is that in each softer band.
In calculating the fit chi-squared,
we must consider both the statistical error $\sigma_{x}$ associated with $x$, 
and that associated with $y$, namely  $\sigma_{y}$. 
Considering error propagation from $x$ to $y$, 
we therefore used a combined quantity 
$\sigma = \sqrt{\sigma_{y}^2 + (a \sigma_{x})^2}$ 
as the overall error to normalize the fit residual at each data point. 
Table 1 shows the parameters, $a$ and $b$, obtained in fitting the CCPs.
In calculating table 2, 
we incorporated a systematic errors of 1.5\%, except for the 0.5--0.8 keV band
in which a larger systematic error of 2\% was assumed to account for the uncertainties in the
CCD contamination described above.


\begin{table*}[t]
 \caption{Parameters obtained by fitting CCPs with equation (2)$^{*}$.}
 \label{all_tbl}
 \begin{center}
  \begin{tabular}{ccccccc}
   \hline\hline
  Energy range  & 0.5--0.8 keV & 0.8--1.0 keV & 1.0--1.2 keV & 1.2--1.4 keV & 1.4--2.0 keV & 2.0--3.0 keV \\

   \hline
  $ a$
                       & $0.14\pm0.02$
                       & $0.26\pm0.02$
                       & $0.37\pm0.03$
                       & $0.40\pm0.03$
                       & $0.86\pm0.05$
                       & $0.54\pm0.04$\\

  $b $
                        & $0.18\pm0.04$
                        & $0.20\pm0.05$
                        & $0.21\pm0.06$
                        & $0.14\pm0.06$
                        & $0.13\pm 0.11$
                        & $0.04\pm0.07$\\[1.5ex]
                        
  $\chi^2 /$d.o.f.
				& 26.34 / 17
				& 27.62 / 17
				& 27.68 / 17
				& 27.73 / 17
				& 25.95 / 17
				& 16.03  / 17     \\                                  
      \hline\hline
      
  \end{tabular}
 \end{center} 

{\small
  \footnotemark[$*$] Systematic errors of 2.5\% is included in the 0.5--0.8 keV band, and 1.5\% in the other band.}
         
\end{table*}


\begin{figure*}[t]
 \begin{center}
   \FigureFile(80mm,80mm)
    {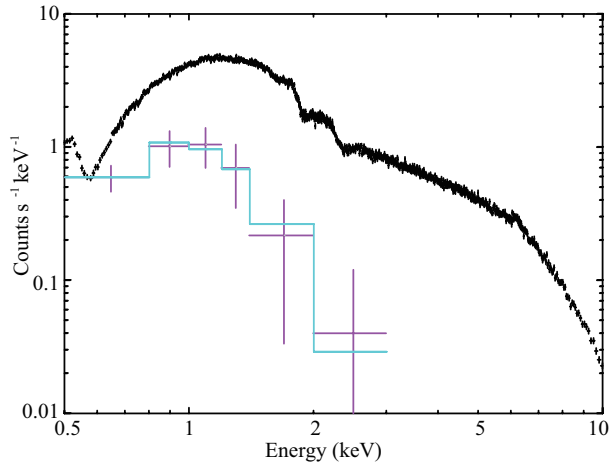}
 \end{center}
\setlength{\belowcaptionskip}{0mm}
 \caption
{Background-subtracted time-averaged spectrum of Mrk 509, obtained by XIS 0 + 3 (black).
Purple crosses show the values of $b$ in table 1, divided by the band width.
Cyan line represents the best-fit {\tt comptt} model with a coronal temperature of 
$T_{\rm e} = 0.45$ keV, and an optical depth of $\tau=16.5$.}
\label{fig:ccp}
\end{figure*}

Since the values of $b$ in table 1
is all non-zero and positive, we divided them by corresponding band width,
and plotted the results in figure 4 in purple  superposed on the time-averaged XIS 0 + 3
spectrum.
Thus, the 6-band CCP analysis suggests the presence of a non-varying, 
softer spectral component in energies below 3 keV.
To examine if the purple data points in figure 4 are well defined,
we carried out the same analysis, changing the fiducial energy range in figure 3 
from 3--10 keV to 4--10 keV or 5--10 keV.
As a result,
the values of $b$ were confirmed to differ by no less than 5\% among the three trials.
Therefore, the purple spectrum in figure 4 can be considered robust.

\subsection{Model fitting to the stable soft component }

To examine properties of the ``stable soft component'' extracted in the previous subsection, 
we tried quantifying its spectrum (purple in figure 4) using typical empirical models.
When the absorption is fixed to $N_{\rm H} = 4.4 \times 10^{20}$ cm$^{-2}$ 
(the Galactic line-of-sight column),
it can be reproduced by a PL model with a photon index of $\Gamma=4.08^{+0.96}_{-0.79}$
($\chi^2/$d.o.f.$ =0.31 / 4$),
or by an optically-thin thermal plasma emission model {\tt apec} 
with a temperature of $0.46\pm0.16$ keV and an abundance  fixed at  0.5 solar
($\chi^2/$d.o.f.$ =3.33 / 4$),
or a blackbody of temperature  $0.16^{+0.04}_{-0.03}$ keV
($\chi^2/$d.o.f.$ =1.35 / 4$).
Therefore, this component is too soft (in terms of the PL modeling) for any leak-through 
signals of the dominant  $\Gamma = 1.8$ PL.
The blackbody  temperature is too hot (in terms of the black body modeling) 
for emission from an optically-thick accretion disk, 
because a standard disk temperature can be calculated to be $\sim 20$ eV 
when the black hole mass is $1.4\times10^{8}M_{\odot}$ and 
the accretion rate is 0.1 times the Eddington rate (section 1).
The plasma emission model is not likely either, as we later find (in subsection 4.1),
slow variations in this soft component.
Furthermore, the lack of particular spectral emission feature at 0.7--1.0 keV
in figure 4 (or similarly in figure 5) rules out the blurred Fe-L emission in interpretation
of this soft component.

As an alternative attempt,
we adopted the MZC standpoint,
and fitted the same soft excess spectrum (purple in figure 4)
by a thermal Comptonization model, {\tt comptt} in {\tt xspec}.
Its seed photon temperature was fixed at 20 eV,  
while its coronal temperature and optical depth were left free.
As a result,
the soft spectrum was reproduced successfully ($\chi^2$/d.o.f.$ = 0.20/3$),
with a coronal temperature of $T_{\rm e} = 0.45^{+45.57}_{-0.31}$~keV,
and an optical depth of $\tau = 16.5~(> 4.6)$.
Since these parameters are reasonable for a Comptonizing corona 
around a BH (cf. Makishima et al. 2008), 
the extracted soft component can be explained in an  MZC viewpoint.
In figure 4, we superpose the best-fit {\tt comptt} model in cyan.
It is nearly identical to the best-fit PL model.

\subsection{Analysis of wide-band spectra}
\label{sec:spectrum}

\begin{figure*}[t]
 \begin{center}
   \FigureFile(140mm,140mm)
    {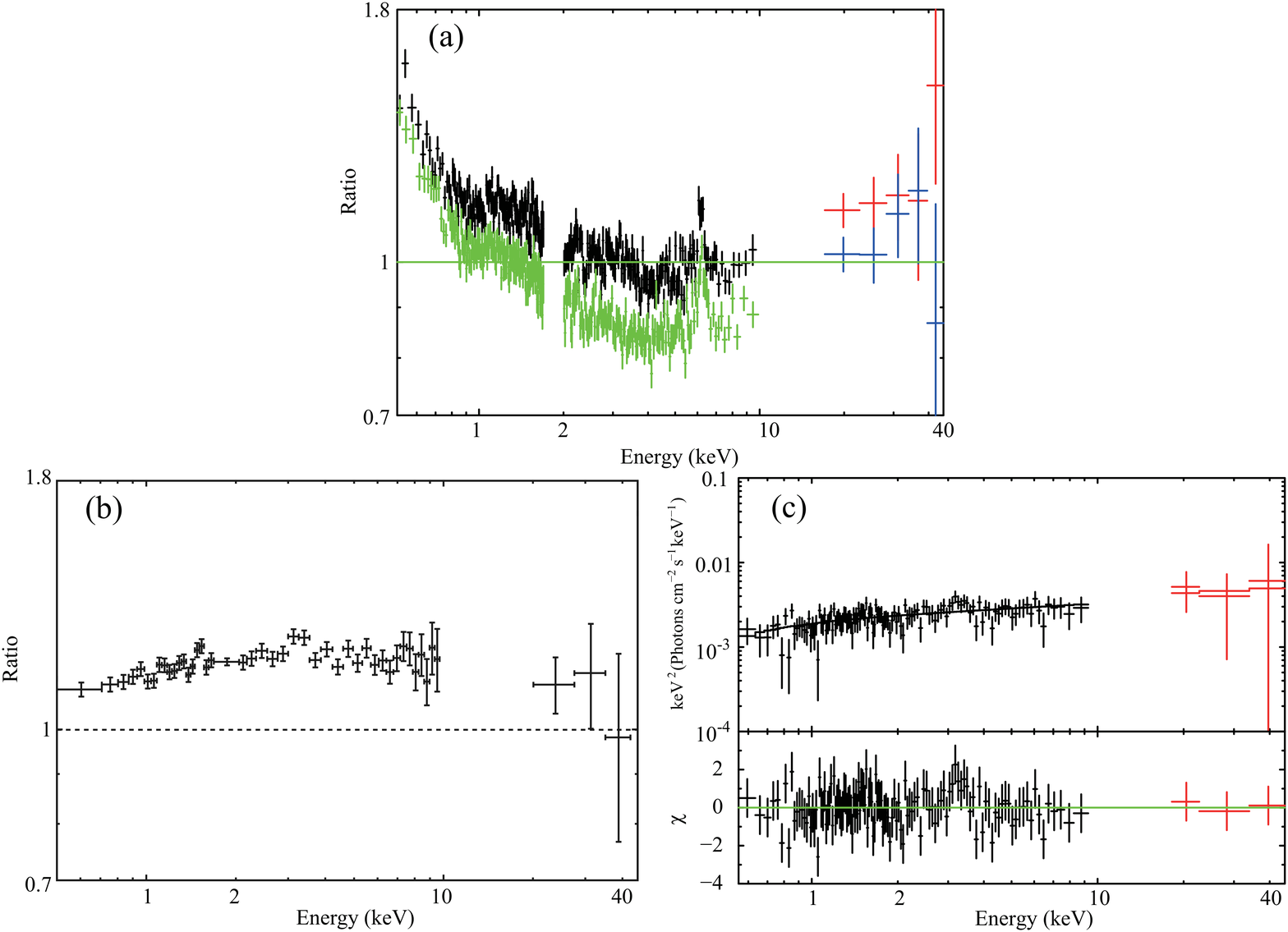}
 \end{center}
 \caption
{(a) Background-subtracted High-phase (black and red) and Low-phase 
(green and blue) spectra,
shown as ratios to a common PL of  photon index  1.8.
(b) Ratio of the High- to Low-phase  spectra.
(c) Difference spectrum in $\nu F_{\nu}$ form obtained 
by subtracting the  Low-phase spectrum from that in the High phase,
fitted with a \texttt{wabs * power} model.}

\label{fig:lcs}
\end{figure*}

\begin{figure*}[t]
 \begin{center}
   \FigureFile(90mm,90mm)
    {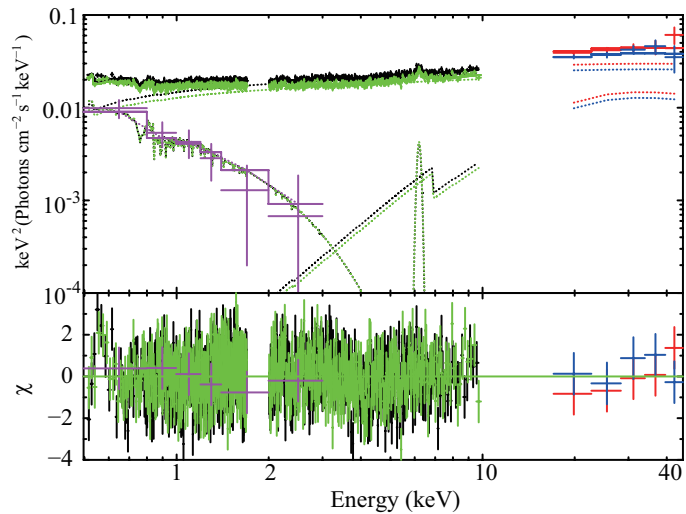}
 \end{center}
 \caption
{Simultaneous fit to the  High- and Low-phase spectra with 
\texttt{wabs * (cutoffpl + pexrav + zgauss + warmabs * comptt)}.
The normalization of \texttt{cutoffpl} and \texttt{zgauss} are independent between the two sets of spectra, 
while the other parameters are tied.
The purple spectrum derived in figure 4 is also fitted simultaneously
with  \texttt{warmabs * comptt}, of which the parameters (including the normalization) 
are tied to those of the High and Low simultaneous fitting.}
\label{fig:lcs}
\end{figure*}

According to our CCP analysis conducted in subsection 3.1,
the 0.5--10 keV spectrum is likely to be decomposed into the stable soft component (in figure 4),
and the variable one having a constant shape 
(as evidenced by the tight linear correlations seen in figure 3).
However, this suggestion obviously needs reconfirmation through more standard analysis 
of wide-band  spectra.
Below, we conduct such attempts, first in a model independent manner utilizing the time variations, and then through spectral model fits.

The 0.5--10 keV variations by $\pm 10$\%
happened to be such that the intensity was intermediate in
the first $\sim1/3$ of the light curve (figure 1),
followed by a low intensity period, 
and then the source brightened during the last 1/3.
Therefore, we divided the entire observation into three periods,
0--60 ks, 60--130 ks, and 130--190 ks, 
and named them Middle, Low, and High phases, respectively.
Then, XIS and HXD-PIN spectra were extracted from the individual phases.
The derived High-phase and Low-phase spectra are shown in figure 5(a),
in the form of ratios to a common power-law with  a photon index of $\Gamma=1.8$.
Thus,
the two spectra both  exhibit a clear soft excess,
as a  flux increase towards the lower band relative to the $\Gamma=1.8$ PL.
Similarly, the HXD data in figure 5(a) reveal a ``hard-excess" effect.
A clear Fe-K emission line is also seen.

Figure 5(b) shows ratios of the High-phase to Low-phase spectra.
Thus, the ratio decreases clearly to both ends of the Suzaku band pass,
of which the soft-band decrease is in agreement with the inference derived in section 3.1.
Quantitatively,
the High-to-Low variation amplitude is $\sim20$\% 
at 3--10 keV, while it decreases to $\sim10$\% at energies below $\sim1$ keV.
This agrees with figure 3(a)
because its implication, i.e., comparable constant (y-intercept) and variable signals,
would halve the variation amplitude at $\lesssim1$~keV compared to that in 3--10 keV.

To quantify the spectral shape change from the  Low phase to High phase,
we subtracted the latter spectrum from the former.
As presented in figure 5(c), 
the derived difference spectrum is featureless and
can be reproduced successfully, with $\chi^{2} /$d.o.f.$ = 25.47/35$,
by a $\Gamma = 1.80\pm 0.06$ PL model,
modified with the Galactic column density $N_{\rm H}=4.4\times 10^{20}$~cm$^{-2}$ (section 1).
Unlike the other two panels in figure 5,
this panel (c) is in the $\nu F_{\nu}$ form, and a gradual bending below $\sim2$~keV is due to
the Galactic-column absorption.
This result, combined with figure 5(b), 
implies that the entire 0.5--45 keV variability can be attributed to intensity changes of
this $\Gamma=1.8$ PL emission,
and the soft and hard excess effects,
seen in figure 5(a), are due to the presence therein of stable signals which 
disappeared in the process of subtraction.

According to canonical interpretations of AGN spectra,
the variable $\Gamma = 1.8$ component found above 
can be identified naturally with the dominant PL-like continuum due presumably
to Comptonization,
and the constant signals in the hard band 
its reflection by cold materials in the accretion disk.
Then, the non-varying signals in the softer band, 
which causes the ratio decrease towards soft energies in figure 5(b), 
can be naturally identified with the stable soft component revealed in subsections 3.1 and 3.2.
We further speculate that this spectral component is responsible for 
the soft excess seen in figure 5(a).

In order to examine whether the spectroscopic soft excess seen in figure 5(a) 
can really be identified with the stable soft component in figure 4 found through the timing 
analysis, we simultaneously fitted the High- and Low-phase spectra with a model of 
\texttt{wabs * (cutoffpl + pexrav + zgauss + comptt)}.
Here, {\tt comptt} is meant to reproduce the soft excess, in terms a successful modeling of the purple spectrum in figure 4. 
In the fitting, only normalizations of \texttt{cutoffpl} and \texttt{zgauss} were 
left free and independent between the two phases,
while the other parameters were set the same between them.
The interstellar absorption, \texttt{wabs}, was again fixed 
at $N_{\rm H} = 4.4 \times 10^{20}$ cm$^{-2}$.
We also fitted simultaneously the stable soft spectrum, 
by the \texttt{comptt}  term consisting the overall model, 
with its parameters tied to  those in the simultaneous fit between the High- and Low-phase spectra.
Both $T_{\rm e}$ and $\tau$ of \texttt{comptt} were left free (and common among the 3 spectra), 
while the seed disk temperature was again fixed at 20 eV.
However,
the fit was not acceptable ($\chi^2$/d.o.f.$=1398.66/1136$),
mainly because of large negative residuals in $\lesssim 1$~keV,
which are considered to be caused by ionized absorbers.
Then, we included an ionized absorber model, \texttt{warmabs}, 
and repeated the fitting with an improved model, 
\texttt{wabs * (cutoffpl + pexrav + zgauss + warmabs *  comptt)}. 

As a result of this model improvement,
the three sets of Suzaku spectra 
acruired in 2010 have been reproduced successfully 
($\chi^{2}$/d.o.f.$= 1216.15/1138$).
The result of this fitting is shown in figure 6 and table 1.
The photon index of the PL component,  
$\Gamma = 1.82^{+0.02}_{-0.03}$, is consistent, within 90\% errors, 
with that obtained in the fit to the difference spectrum.
In addition,
the basic parameters such as $\Gamma$, the reflection fraction, 
and the equivalent width of the Fe K$_{\alpha}$ line
are consistent with those reported by Ponti et al. (2009) using the Suzaku data obtained in 2006.
Most importantly, the soft excess in the two broad-band spectra,
and the soft component constructed from the CCP analysis, 
have been explained simultaneously by a 
thermal  Comptonization component   
with a coronal temperature of $T_{\rm e}\sim 0.5$~keV, 
an optical depth of $\tau  \sim 18$,
and a $y$-parameter of $\sim 1.3$.
These parameters are similar to the results obtained by Mehdipour et al. (2011) 
utilizing a multi-wavelength observation of Mrk 509.
The purple spectrum in figure 6 was explained by 
the exponentially cutoff part of this relatively hard Comptonized component.
However, $T_{\rm e}$ and $\tau$ of this component heavily couple to each other,
and are hence subject to large errors.
For example, a different solution, 
with a much higher coronal temperature of $T_{\rm e}=46$ keV 
and a much lower optical depth of $\tau = 0.32$,
gives $\chi^2$/d.o.f. = 1229.77/1138,
which is only slightly worse than the above quoted best-fit solution.
We have thus confirmed that the CCP-derived soft spectrum (purple in figure 4) 
can be identified with the soft excess seen in figure 5(a). 
In other words, 
the MZC interpretation can consistently explain 
both the  time variability and the  0.5--45 keV wide band spectrum of Mrk 509
including the clear soft excess.


\begin{table*}[t]
 \caption{The parameters obtained by jointly fitting the High-phase, Low-phase, 
 and soft-component spectra.$^{*}$}
 \label{all_tbl}
 \begin{center}
  \begin{tabular}{cccc}
   \hline\hline
   Component & Parameter & High & Low\\

   \hline
   \texttt{wabs} &$ N_{\rm H}^{\dagger}$
                          & \multicolumn{2}{c}{0.044 (fix)} \\[1.5ex]
   
    powerlaw     & $\Gamma$%
                      & \multicolumn{2}{c}{$1.82^{+0.02}_{-0.03}$}\\

                    & $N_\mathrm{PL}^{\ddagger}$%
                      & $1.64^{+0.05}_{-0.10}$
           	&$1.42^{+0.13}_{-0.06}$\\[1.5ex]
                       
     reflection  & $f_\mathrm{ref}$%
                          & \multicolumn{2}{c}{$0.8^{+0.2}_{-0.2}$} \\[1.5ex]

 Fe~I~K$\alpha$   & $E_\mathrm{c}^{\S}$
                        & \multicolumn{2}{c}{$6.41^{+0.04}_{-0.03}$}\\
                               & $\sigma$ (keV)
                               &\multicolumn{2}{c}{$0.10^{+0.06}_{-0.03}$}  \\
 
                   & $N_\mathrm{Fe}^{||}$ %
                        & $2.57^{+0.71}_{-0.68}$
                        &$2.71^{+0.84}_{-0.59}$\\                    

                    & $EW$ (eV) %
                       & $40^{+25}_{-30}$
                       &$49^{+55}_{-27}$\\[1.5ex]
 
 \texttt{warmabs}  & log~$N_{\rm H}^{\dagger}$
 			&\multicolumn{2}{c}{$0.35^{+0.15}_{-0.17}$}\\
						
			&log~$\xi$~(erg cm s$^{-1}$)
			&\multicolumn{2}{c}{$2.03^{+0.04}_{-0.02}$}\\[1.5ex]
 
 \texttt{comptt} & $T_{0}$  (keV)
   			&\multicolumn{2}{c}{0.02 (fix)}\\
			
			&$T_{\rm e}$ (keV)
			&\multicolumn{2}{c}{$0.49^{+0.02}_{-0.02}$}\\
			
			&$\tau$ 
			&\multicolumn{2}{c}{$17.6^{+1.4}_{-1.5}$}\\
			
			&$N_{\rm Comp}^{\#}$
			&\multicolumn{2}{c}{$4.16^{+0.48}_{-0.40}$}\\[1.5ex]

   $\chi^{2}$/d.o.f. &   & \multicolumn{2}{c}{1216.15/1138}   \\
      \hline\hline
      
  \end{tabular}
 \end{center}
   
  	{\small
	\footnotemark[$*$] Fitted model is  \texttt{wabs * (cutoffpl + pexrav + zgauss + warmabs * comptt)}. \\
         \footnotemark[$\dagger$] Equivalent hydrogen column density in  $10^{22}$ cm$^{-2}$. \\
         \footnotemark[$\ddagger$] The power-law normalization at 1 keV, in units of $10^{-2}$~photons~keV$^{-1}$~cm$^{-2}$~s$^{-1}$~at 1 keV. \\
         \footnotemark[$\S$] Center energy in keV in the rest frame.\\
         \footnotemark[$||$] The Gaussian normalization in units of 
         $10^{-5}$~photons~keV$^{-1}$~cm$^{-2}$~s$^{-1}$ \\
           \footnotemark[$\#$] The thermal Comptonization component normalization in units of 
         photons~keV$^{-1}$~cm$^{-2}$~s$^{-1}$.
         }

\end{table*}


\begin{figure*}[t]
  \begin{center}
    \FigureFile(110mm,110mm)
     {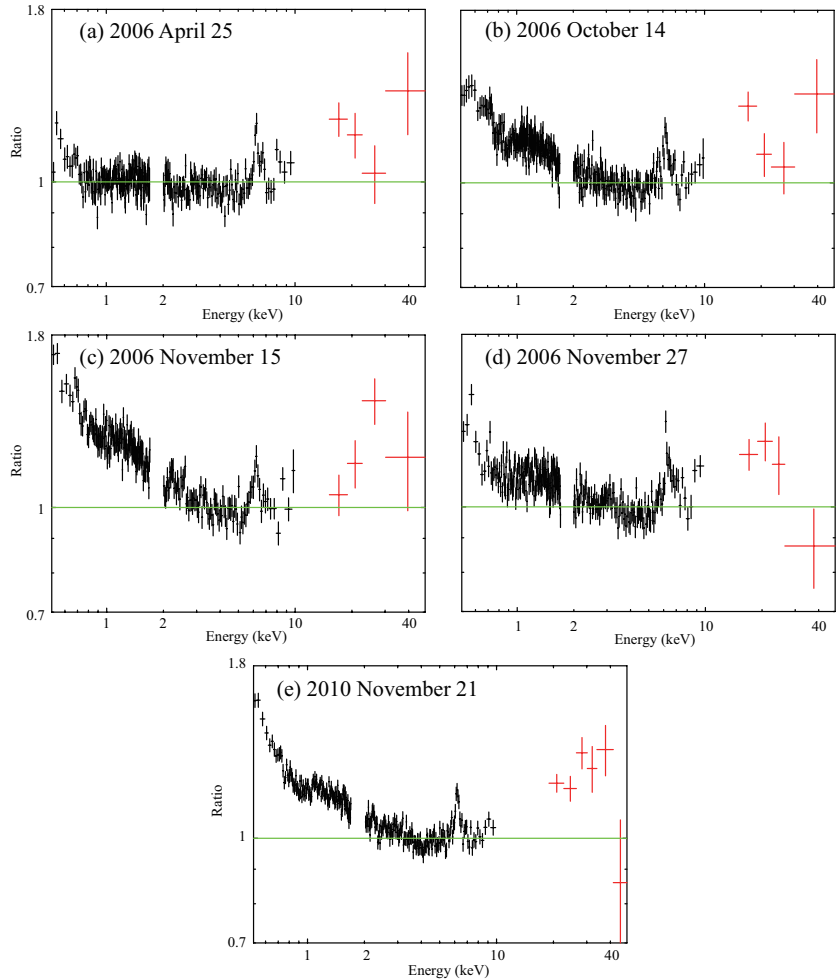}
  \end{center}
  \caption
 {Background-subtracted and time averaged spectra acquired on
 (a) 2006 April 25, (b) 2006 October 14, (c) 2006 November 15, (d) 2006 November 27, and (e) 2010 November 21. All of them are presented as their ratios to a common $\Gamma = 1.8$ PL.
 Panel (e) is similar to figure 5(a), but utilizes the entire exposure.}
 \label{fig:crabratio}
\end{figure*}


\section{Analysis of the Four 2006 Data Sets}
\label{sec:crab}

\subsection{Spectrral ratio analysis}
\label{subsec:fitting}
\begin{figure*}[t]
  \begin{center}
    \FigureFile(130mm,130mm)
     {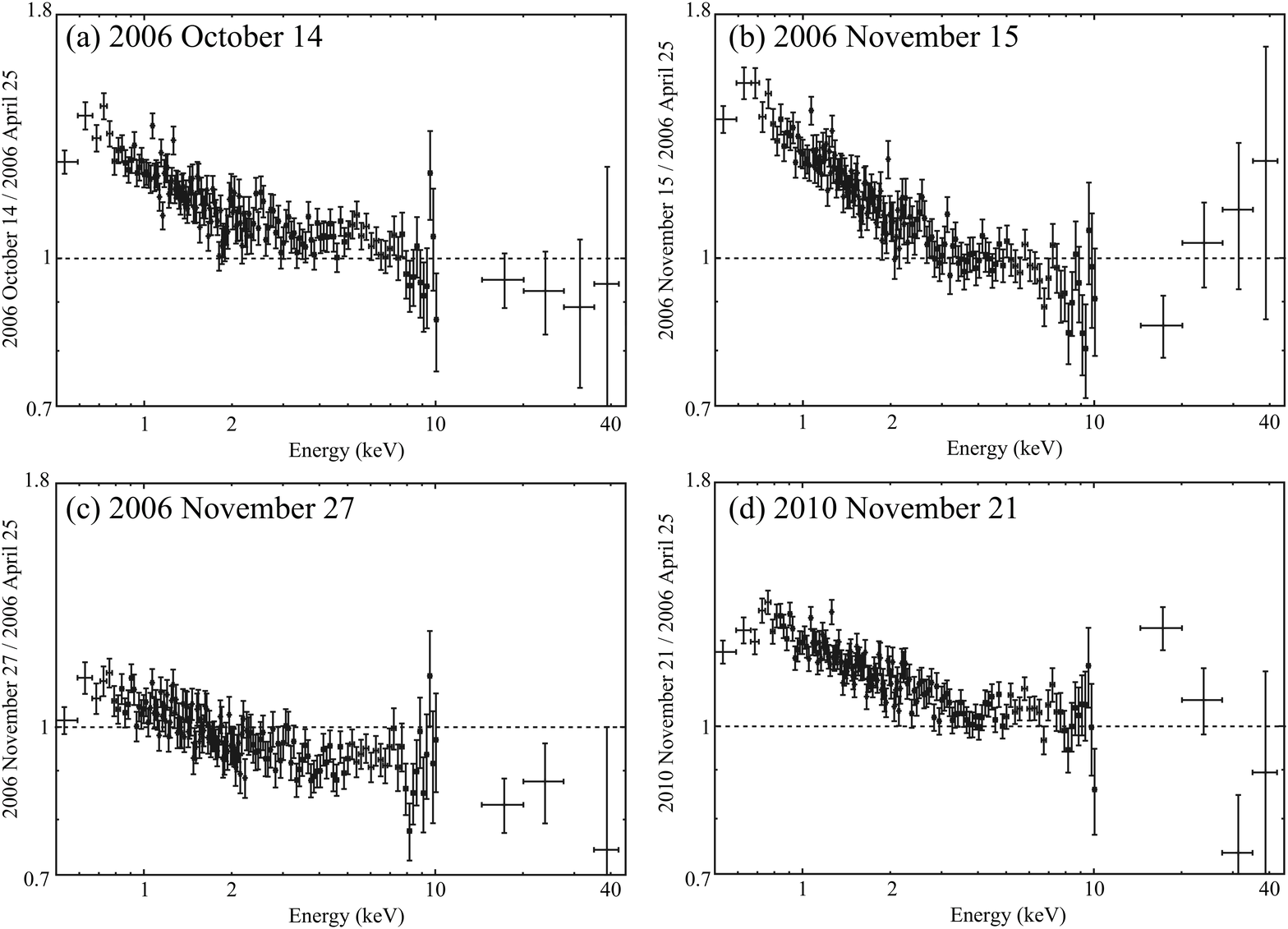}
  \end{center}
  \caption
 {The spectra acquired on (a) 2006 October 14, (b) 2006 November 15,
  (c) 2006 November 27, and (d) 2010 November 21, all divided by that of  2006 April 25.}
 \label{fig:crabratio}
\end{figure*}


The soft excess component in the 2010 November data 
was extracted in section 3  essentially as a component which is  
constant over the 2.2 days of observation.
This raises a suspect that  it could be thermal emission from an extended plasma 
with a temperature of 0.46 keV.
Then, a key issue is whether it varied or not on longer timescales.
This investigation becomes possible when using the four observations
of Mrk 509, conducted in 2006 (section 1, section 2). 
In fact, including the 2010 one, these observations define time intervals of 
2 weeks, 1 month,  6 months, and 4 years between adjacent pointings.   
Therefore, we can study variations of the various spectral components 
on such timescales, 
and obtain valuable information 
which cannot be provided by the single observation in 2010.

To compare shapes of the spectra obtained in different observations,
we made their ratios to a common power-law with a photon index of $\Gamma = 1.8$, 
like in figure 5(a), and show the results in figure 7.
Not only  figure 7(e), but also  figure 7(b-d) for the 2006 October 14, November 15, 
and November 27 reveal steep spectral 
upturns toward lower energies in the 0.5--3 keV band, or the soft excess.
On the other hand, the ratio in figure 7(a)
can be approximated both
by the  $\Gamma = 1.8$ PL from $\sim4$~keV down to $\sim 0.5$ ~keV, 
with only weak indication of the soft excess.
In this way, 
the spectral shapes of Mrk 509 are found to have clearly changed on longer time scales,  
in particular, in the 0.5--3 keV band.

To investigate the spectral variations in details,
we represent in figure 8 
ratios among these spectra, where 
a common denominator is the data obtained on 2006 April 25.
Thus, the four ratios clearly reveal flux changes in energies below $\sim 3$~keV,  
while the ratios in the
3--10 keV band stayed rather constant between 0.9--1.1.
On these long time scales, 
the dominant PL was thus rather stable, 
while significant variability was observed in low energies. 
Very importantly, the soft excess component, 
which appeared in figure 5(a) to be limited to $\lesssim 1$~keV, 
was in fact found to be accompanied by a more gradual continuum 
slope change (figure 7, figure 8) over a wider energy band up to $\sim 3$~keV.
In the 2006 April 25 data (used as a common denominator in figure 8), 
the soft excess was weakest, and was seen only in $\lesssim 0.8$ keV. 
In contrast, 
the soft excess was strongest on 2006 November 15, 
when the continuum was enhanced up to $\sim 3$~keV.
The case of the 2010 data (figure 5, figure 7e, figure 8d) 
is considered intermediate.

\subsection{Wide-band spectral fitting (0.5--45 keV)}
\begin{figure*}[t]
  \begin{center}
    \FigureFile(130mm,130mm)
     {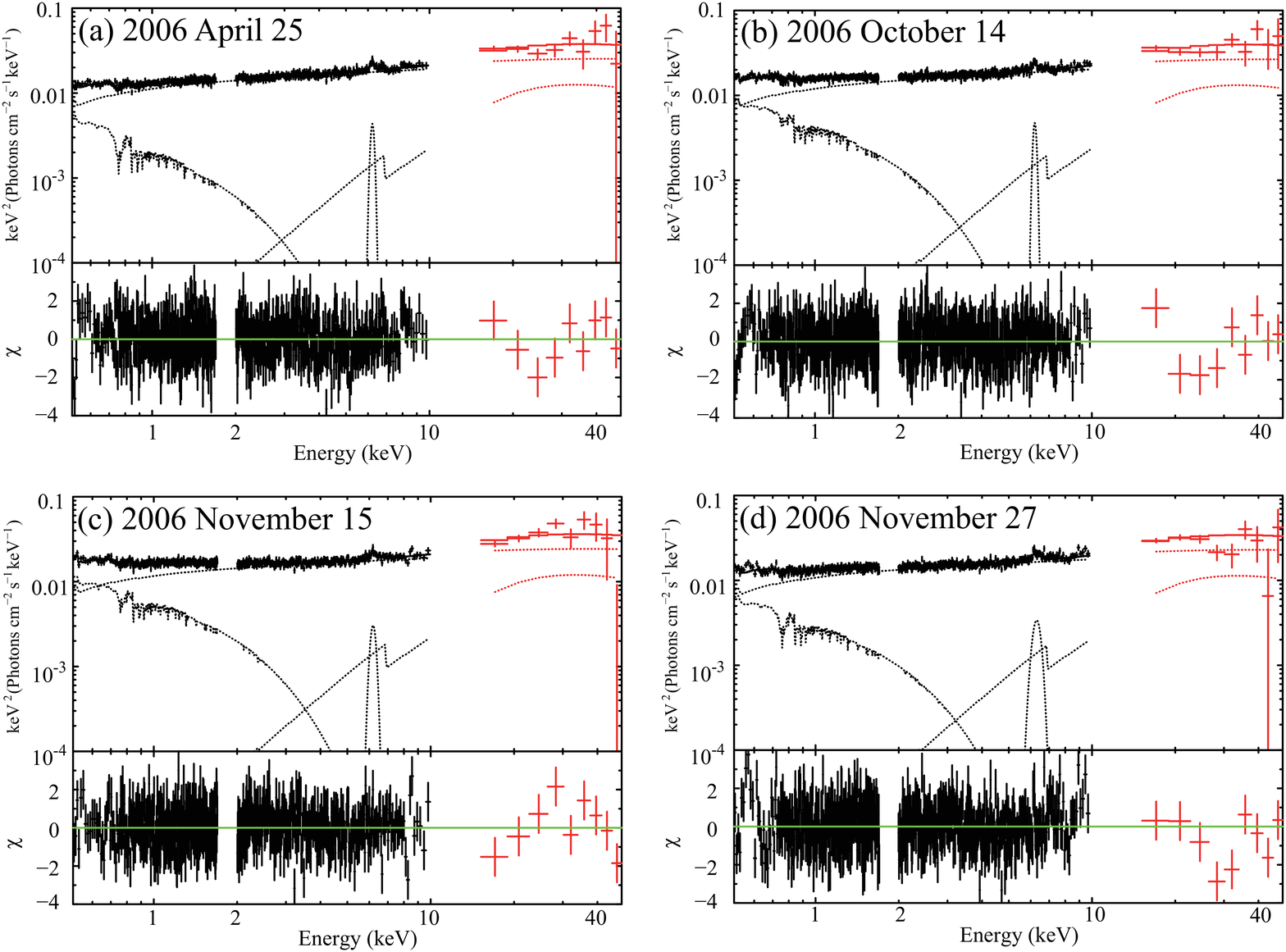}
  \end{center}
  \caption
 {The same as figure 6, but  utilizing the data  obtained in 2006, 
 and without incorporating the soft spectrum in figure 4.}
 \label{fig:crabratio}
\end{figure*}


As a final confirmation of our interpretation presented above, we 
tried to reproduce the four wide-band spectra individually,
by changing  normalizations of the soft component and the PL.
Since this has been successfully conducted on the two 2010 spectra, 
here we fitted the four Suzaku spectra obtained in 2006 with the model,
\texttt{wabs * (cutoffpl + pexrav + zgauss +  warmabs * comptt)}.
In this fit, 
the reflection fraction was fixed at 0.8, which was obtained in  section 3.3, 
and $T_{\rm e}$ of \texttt{comptt} was fixed at the values obtained in section 3.3.

As summarized in figure 9 and table 3, 
the fits were mostly acceptable.
Among the four data sets,
the PL $\Gamma$ was the same within 90\% errors,
and the PL normalization did not change by more than $\pm 10\%$.
Under the assumption of a fixed $T_{\rm e}$,
the  {\tt comptt} normalization varied by a factor of $\sim 1.5$,
as suggested by figure 7,
while $\tau$ by $\sim 10\%$.
Similarly, the ionized absorption changed mildly,
both in the column density and the ionization parameter.
An important fact is 
that the  {\tt warmabs * comptt} components,
found with these 2006 data sets,
are very similar in shape and intensity to that found in the 2010 data (table 2),
even though the  CCP-derived soft component (figure 4) 
was  not incorporated in fitting the 2006 data sets.
Like in the 2010 data (table 2),
the Fe-K line was mostly narrow and centered at 6.4 keV (i.e., neutral fluorescence),
except on 2006 November 27 
when it was slightly ionized and broadened.
Within the rather large errors, 
the Fe-K line equivalent width is consistent 
with being constant over the 5 data sets.  


\begin{table*}[t]
 \caption{The same as table 2, but for the four 2006 data sets.$^{*}$}
 \label{all_parametar}
 \begin{center}
  \begin{tabular}{cccccc}
   \hline\hline
   Component & Parameter & 2006/4/25 & 2006/10/14 & 2006/11/15 &2006/11/27 \\

   \hline
  \texttt{wabs} & $N_{\rm H}$
   		& \multicolumn{4}{c}{$0.044$ (fix)}
   		\\[1.5ex]

   powerlaw     & $\Gamma$%
                       & $1.79^{+0.04}_{-0.05}$
                       & $1.80^{+0.04}_{-0.05}$
                       & $1.82^{+0.06}_{-0.04}$
                       & $1.81^{+0.03}_{-0.04}$\\

                    & $N_\mathrm{PL}$%
                       & $1.25^{+0.09}_{-0.09}$
                       & $1.34^{+0.09}_{-0.09}$ 
                       & $1.30^{+0.12}_{-0.12}$
                       & $1.20^{+0.09}_{-0.05}$\\[1.5ex]
                       
     reflection  & $f_\mathrm{ref}$%
                          & \multicolumn{4}{c}{$0.8$ (fix)}\\[1.5ex]

 Fe~I~K$\alpha$   & $E_\mathrm{c}$

                        & $6.43^{+0.05}_{-0.04}$
                        & $6.43^{+0.05}_{-0.04}$
                        & $6.40^{+0.08}_{-0.09}$
                        & $6.51^{+0.07}_{-0.09}$\\
                                
                                   & $\sigma$~(keV)
                               &$0.10^{+0.09}_{-0.06}$
                               &$0.10^{+0.14}_{-0.08}$
                               &$0.15^{+0.13}_{-0.08}$
                               &$0.23^{+0.12}_{-0.14}$\\

                   & $N_\mathrm{Fe}$ %
                        & $2.89^{+1.00}_{-0.93}$
                        & $2.95^{+1.25}_{-0.92}$ 
                        & $3.02^{+1.66}_{-1.52}$
                        & $4.89^{+2.34}_{-2.23}$\\                    

                    & $EW$ (eV) %
                        & $57^{+45}_{-30}$
                        & $54^{+42}_{-32}$
                        & $59^{+54}_{-30}$
                        & $105^{+60}_{-79}$\\[1.5ex]

\texttt{warmabs} & log~$N_{\rm H}$
   		&$0.34^{+0.15}_{-0.17}$
		&$0.27^{+0.11}_{-0.19}$
		&$0.26^{+0.10}_{-0.18}$
		&$0.27^{+0.12}_{-0.22}$\\
		
           &    log~$\xi$~(erg cm s$^{-1}$)
   		&$2.03^{+0.08}_{-0.03}$
		&$2.08^{+0.04}_{-0.02}$
		&$2.03^{+0.03}_{-0.05}$
		&$2.03^{+0.06}_{-0.03}$\\[1.5ex]

\texttt{comptt}  & $T_{0}$~(keV)
		 & \multicolumn{4}{c}{$0.02$ (fix)}\\
                        
                   &$kT$~(keV)     
                        & \multicolumn{4}{c}{0.49 (fix)} \\      
                        
                     &$\tau$   
                        &$17.0^{+1.6}_{-1.0}$
                        &$18.6^{+0.8}_{-0.7}$
                        &$19.0^{+0.7}_{-0.4}$
                        & $18.4^{+0.8}_{-0.6}$\\   
                        
                      &$N_{\rm Comp}$                                                     
                        & $2.28^{+0.68}_{-0.42}$
                        & $2.41^{+0.49}_{-0.41}$ 
                        & $2.84^{+0.48}_{-0.40}$
                        & $1.90^{+0.40}_{-0.37}$\\

   $\chi^{2}$/d.o.f. &   & 561.52/575  & 647.61/655 & 528.92/557 &  706.26/616\\
      \hline\hline
      
  \end{tabular}
 \end{center}
 {\small
	\footnotemark[$*$] Meanings and units of the symbols are the same as in table 2.}
   
\end{table*}


\section{Discussion and Summary}

\subsection{Summary of the results}

In the present work, we analyzed the five Suzaku datasets of Mrk 509 in four steps.
The first step was a kind of ``dynamic" analysis 
(subsections 3.1, 3.2 and part of subsection 3.3), 
in which we applied the techniques of CCP (figure 2, figure 3), 
and intensity-sorted spectroscopy (figure 5b, figure 5c) to  the 2010 data,
utilizing the $\pm 10\%$ continuum variation.
Then, the presence of a stable soft  spectral component (figure 4)
and a variable $\Gamma \sim 1.8$ PL component (figure 5d) was derived.
The second step was a more ordinary ``static" one (subsection 3.3),
wherein a clear soft excess was noticed
in the 2010 spectra (figure 5a) above the PL continuum.
Combining these steps, 
we showed in the third step (subsection 3.3, figure 6) 
that the stable soft component  derived in the 1st step
can be identified with the soft excess  found in the 2nd step.
Finally, in the 4th step, the analysis of the additional four 
Suzaku data sets acquired in 2006 revealed 
that the  stable soft component was in fact variable
on longer time scales of weeks to months,
without particular correlation to the intensity of the PL continuum.

Through the above analysis procedures, 
the overall spectral and timing behavior of all the five Suzaku data sets
have been explained in a consistent way, 
by a linear combination of the following three spectral components:
\begin{itemize}
\item The ordinary $\Gamma \sim 1.8$ PL, 
   accompanied by a moderate cold reflection with a reflection fraction of $\sim 0.8$.
  The PL intensity varied on a time scale of $\sim 1$ day
   keeping a nearly constant slope,
    while it was not much more variable on the longer time scales.
\item The  soft component with a relatively constant shape,
    which is responsible for the soft excess.
   Its  intensity varied by a factor of $\sim2$ on longer time scales,
   but did not change during the 2010 pointing.
\item The  Fe-K line, which is almost constant and narrow (except on 2006 November 27), 
produced via fluorescence by near neutral matter.
\end{itemize}
In addition, the absorption features due to some ionized absorbers were noticed.
The derived spectral parameters (table 2, table 3) are generally 
consistent with those reported by Ponti et al. (2009).

The soft component in the above scenario has been 
reproduced successful by several alternative models,
including a steep PL with $\Gamma \sim 4.1$,
a blackbody of temperature 0.16 keV,
and a thermal Comptonizaion model based on the MZC viewpoint.
This MZC interpretation is considered reasonable,
because such a scenario has  already been established in Cyg X-1 
(e.g., Makishima et al. 2008),
and the derived value of $y \sim 1.3$ is typical of thermal Comptonization processes.
In addition, the more gradual evolution of the soft component than the PL component
agrees with the MZC properties observed with Suzaku 
from Cyg X-1 (Yamada et al. 2011).

\subsection{Other interpretations}

Is it possible to interpret the soft component  of Mrk 509
in alternative ways, without invoking the MZC view?
As described in section 1,
major explanations so far proposed for the soft excess phenomenon 
in Type I Seyfert galaxies include;
(i) the hardest end of the accretion disk emission, 
(ii) flux leakage through a clumpy absorber,
(iii) thermal emission from some hot plasmas,
(iv) continuum modification due to ionized absorbers,
and
(v) disk reflection dominated by  relativistically blurred Fe-L lines.
As already mentioned in subsection 3.2, (i) can be excluded,
because the equivalent blackbody temperature of the soft excess,
0.16 keV, is much higher than 20 eV,
which is  expected when a BH of $\sim 1 \times 10^8~M_\odot$
is accreting at 10\% of the Eddington limit.
We can also rule out (ii), 
because the equivalent PL slope, $\Gamma \sim 4$,
is too steep for any leakage of the $\Gamma=1.8$ PL.
Similarly, (iii) has been rejected by the long-term changes 
in the soft excess intensity  (section 4).
Below, we therefore examine whether (iv) or (v) can 
explain the soft excess observed from Mrk 509.

The interpretation (iv) assumes a scenario 
wherein absorption  by some ionized absorber 
suppresses the continuum, in intermediate energy ranges
which is above the involved inner-shell edge energies,
while the effect diminishes toward softer energies
because outer-shell electrons are absent due to ionization
(e.g., Schurch \& Done 2008; O'Neil et al. 2007).
This could produce a sort of soft excess as seen in figure 5(a).
In fact, we had to introduce the {\tt warmabs} factor in our fitting
to account for some soft X-ray structures.
However, the difference spectrum shown in figure 5(c) 
bears little evidence of such  local soft X-ray features;
this is difficult to explain under the ionized absorber scenario,
unless some extreme fine tuning is operating 
to control the absorber's column density and ionization degree
in response to the continuum changes.
Furthermore, the simplest case, i.e., no variations in the absorber, 
would predict  $b=0$ in equation (1) and equation (2).
To explain the good linearity of these relations with positive values of $b$,
we would need another (probably independent) fine tuning 
between the continuum and the absorber.
From these considerations,
we conclude that the observed soft excess of Mrk 509
cannot be explained in terms of ionized absorber.

Finally, there remains the explanation (v),
which attributes the soft excess to relativistically broadened 
Fe-L features (e.g., Zoghbi et al. 2009; Nardini et al. 2011).
However, this cannot be applied  to the present observation, 
either, for the following reasons.
First of all, the relatively narrow Fe-K line,
and the moderate strength of disk reflection ($f_{\rm ref} \sim 0.8$),
argues against the presence of strong relativistic effects.
Next, on a time scale of $\sim 100$ ks in the 2010 observation,
the reflection hump intensity (not the fraction) remained constant,
as evidenced by the high-energy drop
of the ratio in figure 5(b), and the lack of such a hump in figure 5(c).
Unless some fine tuning like the ``light bending" 
mechanism (Miniutti et al. 2007) is working,
this suggests that the reflection and the fluorescence lines are
produced at a distance $\gtrsim 100 R_{\rm g}$ from the central BH,
where $R_{\rm g}$ is the gravitational radius 
and corresponds to $\sim 1$ ks for a BH of $10^8~M_\odot$.
Lastly, as revealed by figure 7 and figure 8,
the soft excess in Mrk 509 is an effect 
that spreads over a considerably wider energy range
than is covered by the Fe-L line/edge features.

As discussed for far,
none of the existing explanations, (i)--(v),
succeed in consistently explaining the properties 
of the soft excess observed from Mrk 509.
We hence conclude that our MZC interpretation
provide the most reasonable explanation to the present observation.

\subsection{Implication of the present work}

Through the present study, it has become likely
that the soft excess component seen in Mrk 509 
is produced by a Comptonization process
that has  a $y$-parameter of $y \sim 1.3$. 
This result is consistent with an independent multi-wavelength study on 
Mrk 509 from the optical to the soft X-ray band (Mehdipour et al. 2011), 
which also shows the existence of a thermal Comptonized component 
from $\sim 2$ eV to $\sim10$ keV.
Therefore, the Comptonizing corona in this Seyfert galaxy is 
inferred to be inhomogeneous, namely in a MZC condition,
and is characterized by multiple values of $y$-parameter.
Developed to explain  Cyg X-1 (Makishima et al. 2008, Yamada et al. 2011)
and narrow-line Seyfert galaxies (Marshall et al. 2003),
this MZC view has  thus been applied to a Type I Seyfert, successfully.
In particular, the soft excess component of Mrk 509, revealed in figure 8,
is very similar to the ``soft Compton" component
found in Cyg X-1 (Makishima et al. 2008).
Since the central engine should not differ significantly among Seyfert galaxies,
we expect this view to be applicable to many other Syeferts as well.

The scenario developed here thus decomposed the 0.5--45 keV continuum 
(excluding the hard bump and the Fe-K line) into the two Comptonization component 
(subsection 5.1), the dominant PL and the soft excess, considered to have a larger 
and a smaller values of $y$, respectively. 
It is, however, more natural to consider that the Comptonizing corona in a Seyfert galaxy 
(and of a BHB as well) has its $y$ distributed continuously over a certain range, 
which is represented in the present case by the two discrete values.
Actually, in some other objects, 
we may find an extra spectral component representing higher end of the  $y$ 
distribution. Possible examples include the ``independent hard X-ray compoent" 
found in MCG--6-30-15 (Noda et al. 2011).
The present figure 2(b) is suggestive of a similar effect taking place in Mrk 509. 
In addition to these arguments, 
Yamada et al. (2009) invoked correlated height vs. $\tau$ (or $y$) variations in a Comptonizing corona, 
to explain a Suzaku detection of interesting spectral changes 
in the obscured low-luminosity AGN, NGC 4258.

The present  work is thought to have twofold implications.
On one hand, 
it provides a novel interpretation of the long-standing puzzle of soft excess,
based on a physical analogy to BHBs,
and  in terms of essential properties of the AGN central engine
rather than invoking secondary processes
(e.g., absorption, reflection, or fluorescence).
On the other hand, it shows 
that the AGN continua can actually deviate significantly from a single PL,
like those in BHBs.
Therefore, we need an extreme caution
when trying to model spectral features
(e.g., broad Fe-K lines)
that couple strongly with the continuum.

\end{document}